\documentclass[useAMS,usenatbib]{mn2e}
\usepackage{psfig}
\usepackage{epsfig}
\usepackage{subfigure}

\title[Discovery of three new pulsars in a 610 MHz pulsar survey with the GMRT]{Discovery of three new pulsars in a 610 MHz pulsar survey with the GMRT}

\author[B. C. Joshi et al.]
{B. C. Joshi$^{1}$\thanks{e-mail:bcj@ncra.tifr.res.in}, 
M. A. McLaughlin$^{2}$, A. G. Lyne$^{3}$, 
D. A. Ludovici$^{2}$, \newauthor 
N. A. Pawar$^{1}$, A. J. Faulkner$^{3,4}$, D. R. Lorimer$^{2}$, M. Kramer$^{3,5}$, 
and M. L. Davies$^{4}$\\
$^{1}$National Centre for Radio Astrophysics, Post Bag 3, Ganeshkhind, Pune 411 007, India \\
$^{2}$Department of Physics, West Virginia University, 210 Hodges Hall,
Morgantown, 26506 USA \\
$^{3}$University of Manchester, Jodrell Bank Centre for Astrophysics,
Alan-Turing Building, Oxford Road, Manchester M13~9PL\\
$^{4}$Astrophysics Group, University of Cambridge, Cavendish Laboratory, J J Thomson Avenue, Cambridge CB3~0HE\\
$^{5}$MPI fuer Radioastronomie, Auf dem Huegel 69, 53121 Bonn, Germany}

\begin{document}

\date{\today}

\pagerange{\pageref{firstpage}--\pageref{lastpage}} \pubyear{2007}

\maketitle

\label{firstpage}

\begin{abstract}
We report on the discovery of three new pulsars in the first blind
survey of the north Galactic plane ($45^{\circ} < l < 135^{\circ}$; $
|b| < 1^{\circ}$) with the Giant Meterwave Radio telescope (GMRT) at
an intermediate frequency of 610~MHz. The survey covered 106 square
degrees with a sensitivity of roughly 1~mJy to long-period pulsars 
(pulsars with period longer than 1 s). 
The three new pulsars have periods of 318, 933, and 1056~ms.  Their
timing parameters and flux densities, obtained in follow up
observations with the Lovell Telescope at Jodrell Bank and
the GMRT, are presented. We also report on pulse nulling 
behaviour in one of the newly discovered pulsars, PSR~J2208+5500.
\end{abstract}


\section{Introduction}

The earliest pulsar surveys typically used low frequencies (i.e.~below
$\sim 500$~MHz) due to technological constraints, the steep spectrum
of pulsed radio emission and the larger low-frequency beamwidths
(Large \& Vaughan 1971; Manchester et al. 1978; Hulse \& Taylor 1975;
Stokes et al. 1986; Manchester et al. 1996). Recently, however, most
pulsar surveys have been conducted at frequencies above $\sim1$~GHz to
combat the effects of dispersion and scattering of pulsed signal by
the interstellar medium, and to take advantage of the wider bandwidths
typically available at these frequencies.  The most notable of these
surveys was the Parkes multibeam survey, which has discovered almost
half of the pulsars known to date (see, e.g., Manchester et al.~2001).
The limitation of the small beamwidth at 20~cm in this survey was
overcome by using a 13-beam receiver. This allowed longer (35~min)
integration times, yielding sensitivities of 0.2 mJy to long-period
pulsars with low dispersion measures (DMs), and making this the most
sensitive pulsar survey carried out in the Southern sky to date.

In these surveys with single dish telescopes, there is a trade-off
between the collecting area and the beamwidth, and consequently the
rate of the survey. Such a limitation does not exist for a
multi-element telescope such as Giant Meterwave Radio Telescope
(GMRT), where a large number of smaller antennas can be combined to
provide sensitivity equivalent to a larger dish and yet retain a
relatively large beamwidth.  In this paper, we report on the discovery
of three new pulsars in an ongoing survey of the northern Galactic
plane using this feature of the GMRT at 610~MHz.  The survey was
conducted at this frequency to compromise between pulsars' increased
flux density at low frequencies, interstellar scattering and
dispersion, and beamwidth for the 45-m GMRT antennas.  The
observations, analysis and localization procedure is described in
Sections \ref{oa} and \ref{sa}. The parameters and the integrated
profiles for the new pulsars are presented in Section \ref{res}.

\section[Observations]{Observations}
\label{oa}

\begin{figure*}[h]
\centering
\subfigure{\mbox{\psfig{file=fig1a.eps,width=5.0in}}}\quad
\subfigure{\mbox{\psfig{file=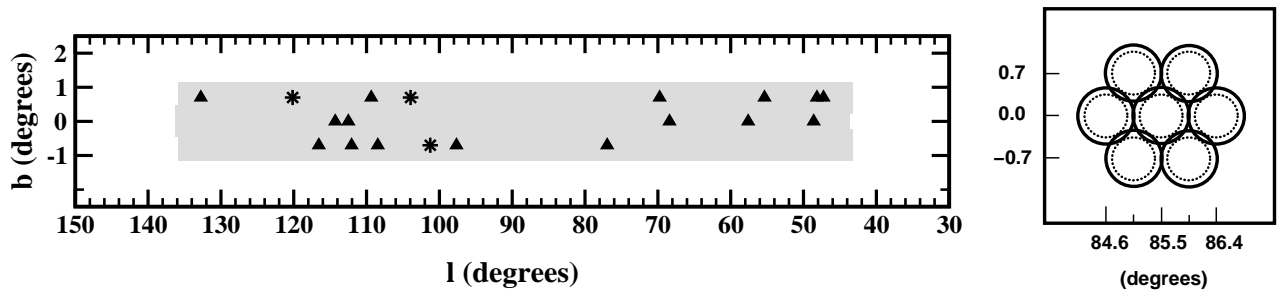,width=1.5in}}}
\caption{The sky coverage of the survey (l = Galactic longitude, b = 
Galactic latitude). The left plot shows the overall 
survey area, where the newly discovered pulsars
  and the known pulsars detected in the survey are marked with stars
  and triangles respectively. The right plot illustrates the arrangement 
of individual pointings for a small fraction of the sky coverage shown 
in the left plot}
\label{fig1}
\end{figure*}

The observations were conducted using the GMRT antennas in an
incoherent array mode at a frequency of 610~MHz. The full width at
half maximum (FWHM) of a 45-m GMRT antenna at this frequency is
0.67$^\circ$ (40 arcmin), giving a sky coverage per pointing of 0.4~square
degrees, about two-thirds of the Parkes 13-beam receiver (Manchester
et al. 2001).  Each antenna provides two hands of circularly polarized
voltages, received using a coaxial waveguide feed with a nominal
bandwidth of 40~MHz. These signals are amplified with GaAs MESFET
based uncooled low noise amplifiers.  The system noise temperature for
this configuration on cold sky is 101 K with a system equivalent
flux density of 316~Jy.  The amplified signals are mixed with
appropriate local oscillators at each antenna to obtain two 32~MHz
signals at an intermediate frequency of 70~MHz, which are transmitted
to the central building using analog modulated optical signals over an
optical fiber. The signals from each antenna are further mixed at the
central building to obtain two 16~MHz sidebands for each polarization
and are digitized with 4~bit precision for processing in an FX digital
correlator.

For our observations, the GMRT correlator was used as a digital
filterbank to obtain 256 spectral channels, each having a bandwidth of
62.5~KHz, across the 16~MHz upper sideband of each polarization
received from each of the 30 antennas of GMRT. The digitized signals
from typically 20 to 25 GMRT antennas were added after detecting the
voltages in each channel, forming an incoherent sum of signals with
the GMRT Array Combiner (GAC). The summed powers in each channel were
then acquired into 16-bit registers every 16 $\mu$s after summing the
two polarizations in a digital backend and were accumulated before
being written to an output buffer to reduce the data volume. The data
were then acquired using a PCI card and recorded to an SDLT tape for
off-line processing. The effective sampling time for this
configuration, used for most of the survey, was 256 $\mu$s.  Roughly 
25 percent of the survey in the beginning was carried out 
with a sampling time of 512 $\mu$s.

High-pass filtering was usually employed in previous surveys, such as
the Parkes multi-beam survey, to minimize the effect of slow gain
variations as well as to remove a mean baseline allowing 1-bit
digitization of data to reduce data volumes.  No such high-pass filter
is present in the entire acquisition chain for our survey.  While this
preserves higher signal-to-noise ratios (S/Ns) in our survey, it also
makes us more sensitive to Radio Frequency Interference (RFI).  Since
high-pass filtering can lead to reduced sensitivity to long period
pulsar candidates, one of the main targets of this survey, we chose
not to filter the data for this experiment.

The survey covered an area of sky defined by Galactic coordinates in
the range 45$^\circ < l < 135^\circ$ and $|b|<1^\circ$.  In
Fig.~\ref{fig1}, we show the 300 pointings comprising the
survey, for a total sky coverage of 106 square degrees.  The survey
pointings were arranged in three strips at constant Galactic latitude
($b = -0.7, 0$ and 0.7) offset with each other by
0.7$^\circ$. Pointings were also separated by 0.9$^\circ$ in
Galactic longitude. Each field was observed for 35 minutes.  In a
typical session, 10$-$15 pointings were observed.  In addition, known
pulsars near the survey region, with flux densities ranging from 1 to
83 mJy at 610 MHz, were observed in some sessions to check the quality
of data.

The observing bandpass can be displayed online and this was
periodically checked to note the presence of any narrowband RFI
feature.  Such features are sometimes present in a small number of
GMRT antennas only and the data quality can be improved by omitting
these antennas at the cost of a marginal reduction of sensitivity. The
antenna selection took this into account whenever possible.

\begin{figure*}
\centering
\psfig{file=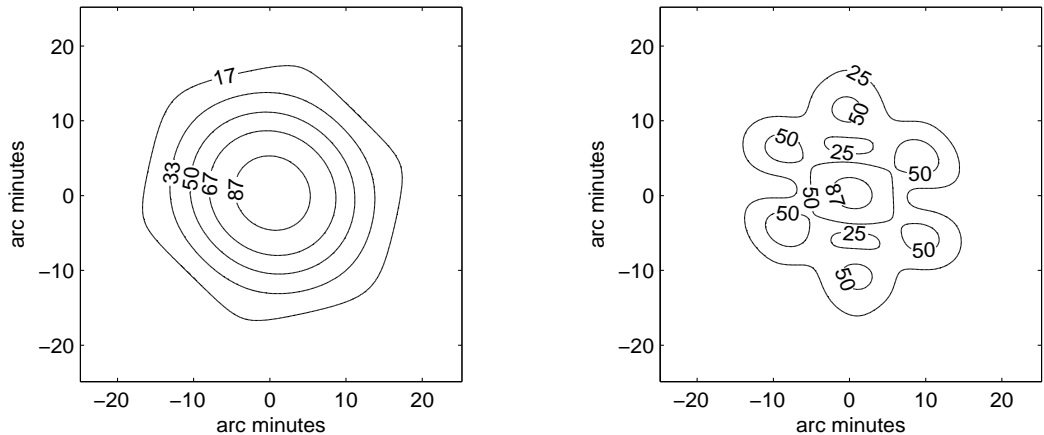,angle=0,width=5.5in}
\caption{The synthesized beams for a phased array of two different 
combinations of the closest GMRT antennas. The left plot shows 
a contour plot for the 
synthesized beam for the closest three antennas C05, C06 and C09, 
whereas the right plot shows the beam for C01, C05, C06, C08 and C09.
The contour levels are labeled by the percentage of the power received 
with respect to the centre of the beam. 
}
\label{fig2}
\end{figure*}

The S/N of the combined output of an incoherent array of more than one antenna 
as compared to that for a single antenna
improves by a factor $\beta \times \sqrt(2 \times N)$, where $N$ is the
number of antennas used and $\beta$ is an experimentally determined
loss factor ($\sim$0.8) introduced while adding antennas incoherently due to 
marginally different power levels across different antennas. 
The configuration, however, retains the 40' FWHM of the 
primary beam of the single antenna. 
An incoherent array of 25 GMRT antennas provides a  S/N five times that of 
a single dish with no change in the resultant FWHM.  
This mode thus provides a faster
coverage of the search region than a single large dish, while
providing an equivalent sensitivity. The 8-$\sigma$
threshold for detecting a pulsed signal with a duty cycle of 10
percent for the configuration used is 0.5 mJy, comparable to the
sensitivity of Parkes multibeam survey when scaled to our lower
observing frequency assuming the mean spectral index for normal 
pulsars to be -1.6 (Kramer et al. 1998). 
 This threshold was calculated using 
the radiometer equation assuming a 
system noise temperature of      91 K  and a sky background 
temperature of 10 K. 
Given the known pulsar detections discussed in
Section 4, we detect on average known pulsars at 50\% of their expected 
S/N, probably due to contamination by RFI. Thus, our true sensitivity is 
likely to be closer to 1~mJy.

\section{Analysis}
\label{sa}

The data were analyzed using the {\sc
  SIGPROC}\footnote{http://sigproc.sourceforge.net} pulsar data
analysis software on computer clusters at Jodrell Bank, West Virginia
University and the National Centre for Radio Astrophysics. The data
were dedispersed using 145 trial DMs ranging from 0 to
2000~pc~cm$^{-3}$, the maximum DMs we expect in the searched region
(Cordes \& Lazio 2003). DMs were spaced more coarsely at higher
frequencies, due to the greater dispersion smearing across each
individual frequency channel.

We searched for periodicities using both a Fast Fourier Transform
(FFT) and a Fast Folding Algorithm (FFA). We summed up to 32 harmonics
in the FFT search and used a threshold S/N of 8. This same threshold
was used in the FFA search, which searched for periodicities greater
than 3 seconds using an algorithm described in Lorimer \& Kramer
(2005).  Known interference periodicities as well as periodicities
occurring in multiple beams were eliminated from our candidate
lists. Pulsar candidates were identified by inspecting the profile over
eight 2-MHz subbands and eight $\sim$4~min time subintegrations in a
composite diagnostic plot. Confirmation observations with the GMRT
were undertaken for any signals which showed the characteristics
expected of a pulsar.

\subsection{Candidate Localization}
\label{local}

The wide beam for the configuration of the GMRT used in the survey
results in a large uncertainty for the pulsar position. However, by
using the GMRT as a phased array, where the phased voltages from
individual antennas are added in the GMRT digital backends, we were
able to obtain refined estimates of the pulsar positions.  The FWHM
for this configuration depends upon the separation between the
different antennas. Simulations were carried out to estimate the
synthesized beam of GMRT for different combinations of GMRT
antennas. Fig.~\ref{fig2} shows the synthesized beams for two
different configurations of GMRT antennas, with FWHM of about 20' and
10'. Three such configurations were selected for localization allowing
a refinement of position upto 3'.

\begin{figure*}
\centering
\psfig{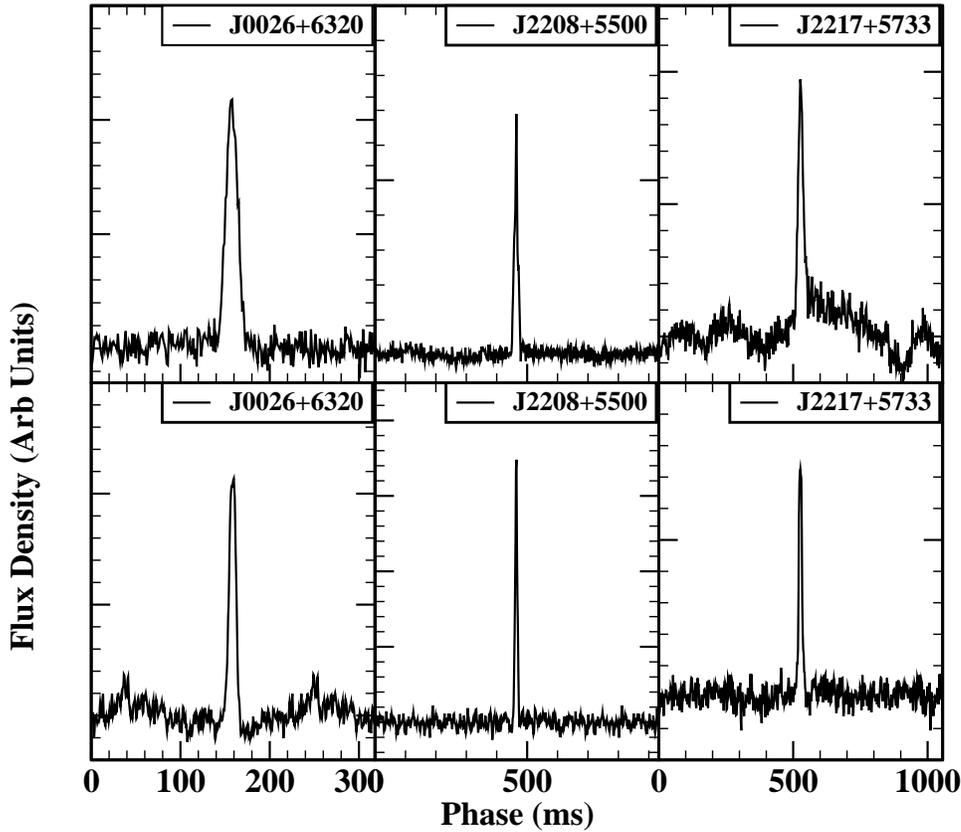}
\caption{Average pulse profiles for PSRs J0026+6320, J2208+5500 and
  J2217+5733. The profiles in the top panels were obtained using GMRT 
  at 626 MHz. The profiles in the bottom panels were obtained using 
  Lovell telescope at 1400 MHz. The features in the baseline of average 
  profiles for PSRs J2217+5733 (626 MHz) and  J0026+6320 (1400 MHz) 
  are probably due to RFI }
\label{figip}
\end{figure*}

\begin{table*}
\caption{Observed and derived parameters of the new pulsars}
\centering
{\begin{tabular}{l|c|c|c}
\hline \hline
Parameter                      &  J0026+6320   &  J2208+5500  & J2217+5733  \\\hline
Right Ascension (J2000)\dotfill&00:26:50.561(8)&22:08:23.72(1)&22:17:55.03(2)\\
Declination (J2000) \dotfill   &63:20:00.87(5) &55:00:08.41(5)&57:33:04.0(2) \\
Galactic longitude (deg)\dotfill&  120.2       &  100.9       &  103.5      \\
Galactic latitude (deg)\dotfill &    0.6       &   $-0.8$       &    0.6       \\
Period (s)\dotfill &0.318357731235(1)  &0.933161074780(6)  &1.05684421944(4) \\
Period Derivative (10$^{-15}$ s/s)\dotfill&0.15006(6)&6.9878(6)&  0.651(2) \\
Epoch (MJD) \dotfill            & 54000.0     & 54315.0     & 54315.0  \\
Dispersion measure (pc cm$^{-3}$)\dotfill&230.31& 101.03      &  162.75 \\
Timing data span (MJD)\dotfill  & 53249$-$54750 & 53864$-$54769&53866$-$54769 \\
Number of TOAs \dotfill          & 147          &      126      &  85 \\
Flux density S$_{626}$ (mJy)\dotfill& 1.9(2)  &  1.4(2)        &  2.2(4)     \\
Flux density S$_{1400}$(mJy)\dotfill& 1.0(3)   &  0.26(8)        & 0.26(8)  \\ 
Pulse width at half maximum W$_{626}$ (ms)\dotfill&15(1) & 14(1) & 26(4)   \\
Pulse width at half maximum W$_{1400}$ (ms)\dotfill&8.1(8)&8.6(6) &13(2)  \\
Spectral Index\dotfill          & $-0.8$        & $-2.09$        & $-2.65$           \\
DM distance (kpc)\dotfill       &  9.7       &  3.5        & 5.0     \\
L$_{1400}$ (mJy kpc$^2$)\dotfill  & 94.8       &  3.2         & 6.5 \\
Inferred surface dipole magnetic Field (10$^{12}$ G)\dotfill& 0.2       & 2.6          & 0.8 \\
Characteristic Age (Myr)\dotfill & 33.6       & 4.2          & 51.5  \\
\hline
\end{tabular}}
\label{tab1}
\end{table*}

\begin{table*}
\begin{center}
\begin{tabular}{lrrrlllllll}
\\
\hline\hline Name & $l$ & $b$ & P & DM & S$_{610}$ & S/N expected & S/N
detected & Distance & Comment &\\
& $\deg$ & $\deg$ &  s  & pc~cm$^{-3}$ & mJy  & & & $\deg$ & &\\
\hline
B2027+37 & 76.9 & --0.7 & 1.216805 & 190.7 & 5.7 &  66 & 46 &
0.07& &\\
B1916+14 & 49.1 & 0.9 & 1.181023 & 27.2 & 1.3 &  16 & * & 0.10 & RFI
&\\
B1915+13 & 48.3 & 0.6 & 0.194631 & 94.5 & 15.0 & 112 & 266 & 0.19
& &\\
B2148+52 & 97.5 & --0.9 & 0.332206 & 148.9 & 7.8 & 57 & 11 &
0.19 & RFI &\\
B2319+60 & 112.1 & --0.6 & 2.256488 & 94.6 & 24.9 & 180 & 66 &
0.22 & RFI &\\
B1952+29 & 65.9 & 0.8 & 0.426677 & 7.9 & 7.0 & 60 & * & 0.23 &
RFI &\\
B2334+61 & 114.3 & 0.2 & 0.495300 & 58.4 & 5.2  & 32 & 12 &
0.23 & RFI &\\
B1929+20 & 55.6 & 0.6 & 0.268217 & 211.2 & 9.9  & 35 & 59 & 0.27
& &\\
B2002+31 & 69.0 & 0.0  & 2.111265 & 234.8 & 7.2  & 74 & 120 & 0.29
& &\\
B1937+21 & 57.5 & --0.3 & 0.001558 & 71.0 & 82.3  & 293 & 122 &
0.30 & RFI &\\
J0215+6218 & 132.6 & 1.0 & 0.548879 & 84.0 & 10.6 & 33 & 9 & 0.37 & RFI &   \\
J2302+6028 & 109.9 & 0.4 & 1.206404 & 156.7 & 6.1 & 18 & 9 & 0.40 & RFI  &\\
B2000+32 & 69.3 & 0.9 & 0.696761 & 142.2 & 3.3 & 10 & 22 & 0.42 & &\\
B2255+58 & 108.8 & --0.6 & 0.368246 & 151.1 & 21.9  & 63 & 74 &
0.43 & &\\
B1919+14 & 49.1 & 0.0 & 0.618183 & 91.6 & 1.9 & 3 & 12 & 0.44
& &\\
B2324+60 & 112.9 & 0.0 & 0.233652 & 122.6 & 10.8 & 22 & 17 & 0.45 & &\\
B1953+29 & 65.8 & 0.4 & 0.006133 & 104.6 & 6.2 & 5 & * & 0.46 &
 &\\
B1914+13 & 47.6 & 0.4 & 0.281842 & 237.0 & 5.5 & 15 & 13 & 0.46
& &\\
B1924+16 & 51.9 & 0.1 & 0.579823 & 176.9 & 4.3 & 5 & * & 0.56 &
Edge of beam &\\
B0154+61 & 130.6 & 0.3 & 2.351745 & 30.2 & 4.3 & 8 & * & 0.52 &
Edge of beam and RFI &\\
B1911+11 & 45.6 & 0.2 & 0.600997 & 100.0 & 2.1 & 1 & * & 0.61 & Edge
of beam &\\
B2351+61 & 116.2 & --0.2 & 0.944784 & 94.7 & 11.3 & 16 & 10 &
0.66 & RFI &\\

\hline
\end{tabular}
\end{center}
\label{tab:params}
\caption[]{Known pulsars with tabulated flux densities that are within the
  searched area of our GMRT survey. Columns give pulsar name, Galactic
  longitude, Galactic latitude, period, DM, 610 MHz flux density, expected
  S/N, detected S/N, distance from beam center, and comments. The 610
  MHz flux density was estimated by scaling from the values in the ATNF pulsar
  database. The expected S/N was estimated given the radiometer
  equation and our observing parameters and was corrected for the
  position of the pulsar in the survey beam (see text). Only known
  pulsars with flux density measurements at two different frequencies are
  included in this table.}
\end{table*}

Each new pulsar was localized by carrying out five pointings using the
phased array, one at the assumed pulsar position and the other four
offset from it by half of FWHM in the north, south, east and west
directions. The pulsar was usually detected in at least two of the
grid pointings. The position of the grid pointing was weighted by the
measured S/N of detection to arrive at a more refined position.  The
localization accuracy was improved by a grid using first the survey
setup with FWHM of 40' and subsequent grids with each of the three
selected phased array configurations, with FWHMs of 20', 10' and
5'. The position of the pulsar was refined after every iteration.  An
experiment with known pulsars of varying flux densities was carried
out to test this procedure. It was observed that this procedure
yielded the final positions with up to 3' accuracy in about 7 hours of
observations for a weak pulsar (Flux density at 626 MHz of about 2 mJy). 
The required observing time for this procedure is less for a pulsar 
with higher flux density.

PSR~J0026+6320 is very close to the beam centre in the discovery
observations, so its position was refined using a grid pointing
procedure similar to that described above using the Lovell telescope
at Jodrell Bank Observatory at a higher observing frequency of 1.4~GHz. The new localization procedure described
here was very useful for the other two pulsars, particularly for PSR
J2217+5733, as these pulsars were discovered at the edge of the beam.

\subsection{Timing observations and analysis}
\label{to}

Once a position accurate to approximately 3' was obtained using the
procedure outlined above, the new pulsar was observed using 76-m
Lovell telescope at Jodrell Bank Observatory on multiple epochs to
obtain a timing solution. The Lovell telescope is equipped with a dual
channel cryogenic receiver with a receiver temperature of about 35
K at 1400 MHz. This receiver is sensitive to two hands of
circular polarizations. The radio frequency signals are down-converted
to an intermediate frequency at the focus of the telescope and brought
to the receiver room, where these are further down-converted
before being fed to the pulsar hardware. The pulsars
were observed with a 32-channel dual hardware filterbank across a 32
MHz bandpass centered around 1400 MHz.  The total intensity data were
then dedispersed in a hardware dedispersion unit and folded
synchronously at the nominal topocentric period of the pulsar for
subintegrations of between 1 to 2 minutes. The folded data for
typically 6 to 10 subintegrations were written to disk for every epoch
and observations were usually made at approximately regular intervals
of between 10 to 15 days.  A time stamp derived from a hydrogen maser
clock was recorded with each folded profile.  The time-span covered by
these observations as well as the total number of times of arrival
used are indicated for each pulsar in Table \ref{tab1}.

Total intensity profiles were obtained by adding the subintegrations.
These were then cross-correlated with a standard template to give
pulse topocentric times of arrival which were then corrected to the
Solar system barycenter using JPL ephemeris DE200 (Standish
1982). Assessment of arrival time residuals, which are the differences
between actual pulse arrival times and times calculated from a simple
model involving an assumed position of the source and its
rotation, provides measurements of pulsar parameters. Specifically, we fit for pulse phase, right ascension,
declination, rotational frequency and its first time derivative. The
measured values of these parameters from the timing analysis are given
in Table \ref{tab1}.

\section{Results}
\label{res}

Three new pulsars have been discovered in the data analyzed so far
(see Fig.~\ref{fig1}).  Their average profiles at 626 MHz, obtained
using the GMRT, and those at 1400 MHz with the Lovell telescope are
presented in Fig.~\ref{figip}.  Measured and derived parameters for
the three new pulsars are presented in Table~\ref{tab1}.  The flux
densities at 626 MHz were estimated by scaling the data by the
expected system noise, while those at 1400 MHz were calibrated against
a noise source.

There is no obvious association of the newly discovered pulsars with
either globular clusters or supernova remnants.  Likewise, these
pulsars are also not associated with any known high-energy source. The
radio luminosities, magnetic fields and characteristic ages of these
pulsars are similar to the normal pulsar population.

We detected 16 out of 22 known pulsars that were in our surveyed area
(see Table~2). To evaluate the sensitivity of our survey in terms of
these detections, the expected S/N for each pulsar was calculated
using the following procedure.  The flux density of the known pulsars
at 610 MHz was estimated using the flux density at 400 and 1400 MHz
and/or the spectral index quoted in the ATNF pulsar catalog
(Manchester et al.~2005). For PSR J0215+6218, 
the flux density at 610 MHz was taken from Lorimer et al. (1998).  This
estimate was then corrected for the position of the pulsar in the
search beam assuming a Gaussian beam with FWHM of 0.67$^\circ$. The
noise was estimated using the radiometer equation (Lorimer and Kramer
2005) assuming 25 GMRT antennas in an incoherent array. The system
temperature took into account the sky brightness temperature for each
field, obtained from an all-sky map at 610 MHz computed using Global
Sky Model (de Oliveira-Costa et al.~2008).  Pulse width
at ten percent as
quoted in the ATNF catalog was used wherever available. Otherwise, the
duty cycle was assumed to be 10 percent. 

\begin{figure}
\psfig{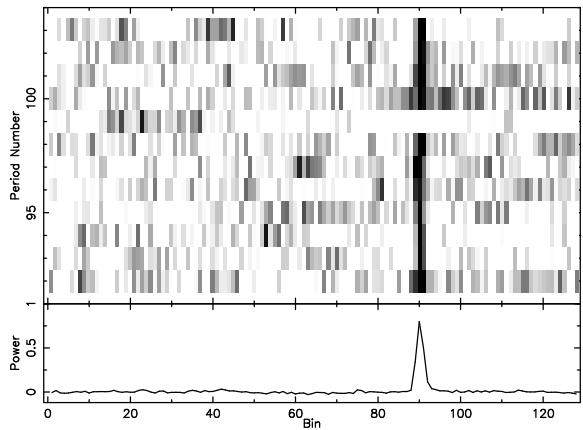}
\caption{The top plot shows the modulation of pulsar energy for 
successive 16 period subintegrations as a contour plot. Lack of emission in 
subintegration 99 in the bins corresponding to the pulse window 
in the average profile, shown in the lower plot, suggests 
nulling for these 16 periods.} 
\label{fig8}
\end{figure}

A comparison of the expected S/N with the detected S/N indicates that 
we detect on average 8 pulsars at 50\% of their expected S/N and  5 pulsars 
with a S/N greater than expected.  Three pulsars were detected with expected 
S/N, while three non-detections were consistent with these pulsars being 
weaker than our expected sensitivity. Modeling the effects of inter-stellar 
scintillations and/or errors in the quoted spectral index of low frequency 
flux densities for the 8 pulsars, detected with lower S/N, 
still leaves a significant discrepancy 
between the expected and detected S/Ns. 
There does not seem to be a correlation between DM, period or
Galactic latitude and S/N loss,
suggesting that propagation effects are not to blame. We therefore 
attribute the loss to RFI, which was considerable in the latter part 
of the survey. All our detections except PSR~B1919+14 had a beam position 
corrected flux density of about 1 mJy, which we adopt as the sensitivity 
of the survey.

Taking account of this empirical sensitivity limit, we used the pulsar
population model from Lorimer et al.~(2006) to predict the likely
yield for our survey. Using the Monte Carlo techniques and the
parameters for model C' described in that paper, we find that our
survey should have detected about 20 pulsars, in excellent
agreement with our total of 16 redetection and 3 new discoveries found
in practice. No new millisecond pulsars were detected in this survey. To
model the expected millisecond pulsar detections, we use the
population models developed by Smits et al.~(2009), and find the
number of millisecond pulsar detections is expected to be of order
unity, i.e.~consistent with our lack of new detections.

The above population estimates do not predict the number of rotating
radio transients that may be present in the survey data.  We are
working on RFI excision algorithms that will allow
us to probe this area of the neutron star population using
single-pulse searches. The results of this study will be presented
elsewhere.
 
\begin{figure}
\psfig{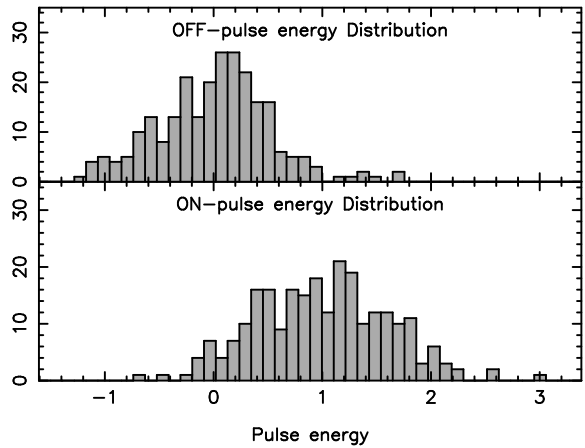}
\caption{The plot shows the distribution of pulse energies in the 
OFF-pulse (top panel) and ON-pulse (bottom panel) windows normalized 
by mean ON-pulse energy. See text for the 
definition of ON-pulse and OFF-pulse windows. The abscissa is dimensionless and 
is labeled in multiples of mean ON-pulse energy. The excess at zero 
energy in the ON-pulse energy distribution indicates the fraction of null subintegrations.}
\label{fig9}
\end{figure}

Follow up observations of PSR J2208+5500 with 13 antennas of GMRT in a
phased array mode (145-m equivalent single dish) at 626 MHz indicate a
nulling behaviour similar to other nulling pulsars. The phased array data were
dedispersed to the nominal DM of the pulsar and were folded every 16
periods, with 128 bins across the pulse. As an example, the
pulse energy modulation in 16 period subintegrations, for a  
duration of 180 s 
out  
of a 4000 s observations, is 
shown in Fig.~\ref{fig8} alongwith the average profile of the pulsar. 
In this plot, the lack of any significant emission is
evident at subintegration 99. This was investigated further by using
the following procedure. First, a baseline, estimated using bins 9 to
69 away from the pulse, was subtracted from the data for each
subintegration. Then, two sequences were formed by averaging the
energy in bins 109 to 114 (OFF-pulse energy) and 87 to 92 (ON-pulse
energy). The two sequences were then compensated for interstellar
scintillation by the mean ON-energy for a block of 300 periods.  The
procedure is similar to that used for detecting pulse nulling in
single pulse sequences (Ritchings 1976; Vivekanand 1995).

A visual examination of the two sequences for the entire 4000 s 
of data shows sudden drop of pulse
energy for 13 subintegrations (e.g.~similar to subintegration 99) similar to
pulse nulling. The average profile for these nulled 
subintegrations shows no detectable emission. 
The average energy  in the pulsed emission during burst 
subintegrations ({\it i.e.} subintegrations with detectable 
emission in the ON-pulse window) is about 18 times higher 
than that during the nulled subintegrations.

The energy in the scaled sequences were also binned to
40 bins to form ON-pulse and OFF-pulse energy distributions shown in
Fig.~\ref{fig9}.  The ON-pulse histogram in Fig.~\ref{fig9} 
shows an excess at zero energies indicating fraction of nulled 16
period subintegrations. This fraction, called the nulling fraction, can be
estimated by removing a scaled version of OFF-pulse histogram at zero
energy from the ON-pulse histogram. 
The procedure used above is not sensitive to nulls shorter than 16 periods.
Thus, a lower limit of 7.5 percent nulling fraction is estimated from 
the distribution presented in Fig.~\ref{fig9}.   
An examination of pulse sequences averaged over fewer numbers of periods 
confirms the nulls mentioned above, but indicates more frequent shorter 
nulls. However, the signal to noise ratio for shorter subintegration 
is not sufficient to estimate the nulling fraction more accurately. 
Higher sensitivity observations at lower frequencies will be useful 
to characterize nulling in this pulsar in the future.

The nulling behaviour for PSR J2208+5500 is similar to other 
nulling pulsars, such as PSRs B0809+74 or B0834+06. Both these pulsars 
have small nulling fractions and periods similar to PSR J2208+5500. The  
pulsed emission of PSR J2208+5500 seem to occur in long bursts interspersed 
with short 1 to 16 period nulls similar to PSR B0834+06 (Ritchings 
1976). In particular, the 
distributions in Fig. \ref{fig9}, apart from an overlap between ON-pulse 
and OFF-pulse distributions,  are very similar in shape to those for 
PSRs B0809+74 and B0834+06. The factor 
by which the radio emission decreases during null ($\eta$) is 
much less than that reported for pulsars such as PSR B0031-07, B0809+74 
and B1944+17 (Lyne \& Ashworth 1983; Deich et al. 1986; Vivekanand \& Joshi 1997), but is still significant. Higher sensitivity observations at a lower 
frequency are required to characterize pulse-nulling in 
terms of these parameters.

\section*{Acknowledgments}
The Giant Meterwave Radio Telescope is operated by 
National Centre for Radio Astrophysics, which is funded by Tata 
Institute of Fundamental Research and Department of Atomic Energy. 
BCJ acknowledges the use of 72-node cluster on the NCRA East Campus, 
funded by Department of Atomic Energy. 
MAM is an Alfred P. Sloan research fellow.
MAM and DRL acknowledge support from West Virginia EPSCoR in the
form of a Research Challenge Grant.

\label{lastpage}

\end{document}